\documentclass[%
 reprint,
showkeys,
 amsmath,amssymb,
 aps,
floatfix,
]{revtex4-2}
\usepackage{xcolor}
\usepackage{graphicx}% Include figure files
\usepackage{dcolumn}% Align table columns on decimal point
\usepackage{bm}% bold math
\usepackage{hyperref}% add hypertext capabilities
\usepackage{color}
\usepackage{gensymb}
\usepackage{siunitx}
\usepackage{array}
\usepackage{graphicx}
\begin{document}

\preprint{APS/123-QED}

\title{Consistent evaluation of continuum scale properties of Graphene}% Force line breaks with \\

\author{Sourabh S Gandhi}
 \affiliation{Department of Civil Engineering, Indian Institute of Technology Kharagpur, West Bengal, India - 721302.}%Lines break automatically or can be forced with \\
\author{Puneet Kumar Patra}%
\email{Corresponding Author: puneet.patra@civil.iitkgp.ac.in}
\affiliation{%
Department of Civil Engineering and Center for Theoretical Studies, Indian Institute of Technology Kharagpur, West Bengal, India - 721302}%

\date{\today}% It is always \today, today,

\begin{abstract}
We handshake statistical mechanics with continuum mechanics to develop a methodology for consistent evaluation of the continuum scale properties of graphene. The scope is kept limited to elastic modulus, $E$, which has been reported to vary between 0.912 TPa to 7 TPa, Poisson's ratio, $\nu$, which has been reported to vary from being negative to a value as large as 0.46, and effective thickness, $q$, whose value varies between 0.75 \AA to 3.41 \AA.  Such a large scatter arises due to inconsistent evaluation of these properties and making assumptions that may not be valid at atomistic scales. Our methodology combines three separate methods -- uniaxial tension, equibiaxial tension, and flexural out-of-plane free vibrations of simply supported sheets, which, when used in tandem in MD, can provide consistent values of $E, \nu$ and $q$. The only assumption made in the present study is the validity of the continuum scale thin plate vibration equation to represent the free vibrations of a long graphene sheet. Our results suggest that -- (i) graphene is auxetic with its Poisson's ratio increasing with increasing temperature, (ii) with increasing temperature, $E$ decreases, and (iii) the effective thickness increases with temperature.
\end{abstract}

\keywords{Elastic properties, Graphene, Normal Modes, Vibration, Molecular Dynamics}%Use showkeys class option if keyword
\maketitle

\section{INTRODUCTION}
With the advances in technology, a new class of materials called two-dimensional materials has been created, wherein electrons are free to move in a two-dimensional plane, but their out-of-plane displacement is severely restricted. Some examples of two-dimensional materials are -- graphene, graphane, graphyne, borophene, silicene, etc. Of these, possibly the most important material is graphene. It comprises a single layer of $sp^2$ hybridized carbon atoms, arranged in a regular hexagonal pattern \cite{novoselov2004electric}. Several important allotropes of carbon, such as graphite, carbon nanotubes, fullerene, etc. may be obtained from graphene. It exhibits some of the best-known electrical\cite{pietronero1980electrical}, chemical \cite{del2018adsorption}, thermal \cite{ghosh2008extremely}, and mechanical \cite{lee2008measurement} properties. For example, the electrical conductivity of graphene is at least three times that of Copper \cite{pietronero1980electrical}, and its thermal conductivity of $3080 – 5150$ W/m-K \cite{ghosh2008extremely} is almost an order of magnitude higher than Copper. Its very high surface area $\sim 2630$ m$^2$ / g makes it a good candidate in applications related to chemical adsorption \cite{szczkesniak2017gas}. It has a breaking strength of $\sim 130$ GPa\cite{lee2008measurement} while bearing an extension up to $25\%$\cite{lee2008measurement}. These extra-ordinary properties make graphene a very attractive material for use in several applications such as composite materials\cite{stankovich2006graphene}, electro-mechanical resonators \cite{bunch2007electromechanical}, strain sensors \cite{hosseinzadeh2018graphene,fu2018high,raju2014wide}, nano-composites\cite{potts2011graphene,papageorgiou2015graphene}, etc. moduli

The use of graphene in potential continuum scale applications, where graphene acts as a reinforcing agent, relies on the accurate knowledge of the continuum scale mechanical properties such as elastic modulus ($E$), shear modulus ($G$et al.), Poisson's ratio ($\nu$), effective thickness ($q$), etc. Information on these mechanical properties is typically required \textit{apriori} for performing initial calculations or for computing the initial strength of graphene-coated materials. They serve as inputs in the ``rule of mixture''. For example, the effective elastic modulus of a graphene-based composite is given by: 
\begin{equation}
    E_{eff}=E_G \times V_{G} + E_M \times V_{M},
    \label{eq:E_eff}
\end{equation} 
where $E_G$ ($V_G$) and $E_M$ ($V_M$) are the elastic moduli (volume fraction) of graphene and matrix, respectively. Customarily, $V_M \approx 98\%-99.5\%$ while $E_M \approx 2.0-3.0$ GPa \cite{wu2014effect,young2018mechanics}. Preliminary calculations show that $E_{eff}$ is significantly dependent on $E_G$. For example, with $E_G = 1$ TPa, $E_M = 2$ GPa, $V_G = 0.5\%$ and $V_M = 99.5\%$, $E_{eff} \approx 7$ GPa. On the other hand, if $E_G$ is taken as 3.84 TPa, as has been reported in the literature, $E_{eff}$ changes to 21.2 GPa. The two composites, although made from the same materials, will have entirely different responses in the linear regime. Similarly, the effective shear modulus depends significantly on the shear modulus of graphene. Likewise, if the thickness of graphene is changed, the longitudinal and transverse moduli of a functionally graded nanocomposite change significantly as per the Halpin-Tsai model \cite{ghafaar2006application}. Further, an accurate estimate of Poisson's ratio of graphene is necessary to evaluate the performance of the nanocomposites in the context of stress concentration\cite{yu2008influence}, buckling response \cite{ellul2009effect,javani2020thermal}, vibration response \cite{zhang2020vibration,wang2019two}, etc.  

\begin{table*}[htbp]
\centering
 \begin{tabular}{| c | c | c | c | c |} 
 \hline
 \textbf{Authors} & \textbf{Method} & $E$ (TPa) & $\nu$ & $q$ (\AA) \\
 \hline
 Arghavan et al.\cite{arghavan2011free} & MD and FEM & 1.0(inplane), 0.11(flexure) & 0.16 & 3.4 \\ 
 \hline
 Wang et al.\cite{wang2014mechanical} & MD & 1.034 & -- & 3.35 \\
 \hline
 Zhao et al.\cite{zhao2013mechanical} & MD & 0.856(ZigZag), 0.964(Armchair) & 0.143/0.157 & 3.41 \\
 \hline
 Kalosakas et al.\cite{kalosakas2013plane} & MD & 1.0 & 0.22 & 3.35 \\
 \hline
 Tsai et al.\cite{tsai2010characterizing} & MD and FEM & 0.912 & 0.261 & 3.35 \\
 \hline
 Thomas et al.\cite{thomas2018assessment} & MD and FEM & 0.939 & 0.223 & 3.34 \\
 \hline
 Kam et al.\cite{kam2013graphene} & MD and FEM & 3.84 & 0.32 & 0.87 \\
 \hline
 Oded et al.\cite{hod2009electromechanical} & DFT & 7.0 & -- & 0.75 \\
 \hline
 Shao et al.\cite{shao2012temperature} & DFT & 1.17/1.2 & -- & 3.35 \\
 \hline
 Zhou et al.\cite{zhou2013elastic} & Molecular Mechanics & $381-385 (N/m)$ & 0.42-0.46 & -- \\
 \hline
\end{tabular}
\caption{The values of $E,\nu$, and $q$ as reported by different authors. Here MD = molecular dynamics, FEM = finite element method, and DFT = density functional theory. Notice the large scatter in the reported values. }
\label{tab:tab1}
\end{table*}

In view of the importance of the mechanical properties of graphene, researchers have made (and are still making) numerous efforts in determining the correct effective continuum scale properties of graphene. Such efforts have revolved around both experimental and numerical techniques, some of which we highlight next. Using atomic force microscopy, Lee et al.\cite{lee2008measurement} conducted nano-indentation tests on graphene flakes of $1-1.5 \mu$m diameter and found $E = 1$ TPa under the \textit{assumption} that $\nu = 0.165$ and $q=3.35$ \AA i.e. they are the same as that in graphite. Ryan et al.\cite{nicholl2015effect}, on the other hand, used interferometry to deform both free-standing and restrained graphene flakes, and compared the deflected shape with the bulge test equation to obtain an estimate of thickness scaled elastic modulus: $E \times q = 340$ N/m, under the \textit{assumption} that $\nu = 0.165$. Antonio et al.\cite{politano2015probing} performed a phonon dispersion based experiment on graphene deposited on metallic surfaces and reported $E \times q = 342 N/m$ and $\nu = 0.19$, without commenting anything on $q$. Notice that in all experimental techniques, one or more variables have been assumed. 

Similar assumptions also feature in the numerical determination of mechanical properties. 
Under the assumption that $q=3.4$ \AA, Liu et al.\cite{liu2007ab} used the ab-initio method for assessing the phonon instability in graphene at 0K and obtained $E=1.05$ TPa and $\nu = 0.186$. Assuming the $q$ and $\nu$ of graphene to be the same as that of graphite, Jin et al.\cite{jiang2009young} performed large scale constant temperature MD simulations on graphene using Brenner-II potential and equated the resulting standard deviation of displacement of the atoms with the equation derived by Krishnan et al.\cite{krishnan1998young} to obtain $E=0.9-1.1$ TPa. Kim et al.\cite{kim2011effective} performed classical molecular dynamics (MD) simulations for graphene to study the flexural wave propagation in them. By comparing the dispersion characteristics obtained from MD simulations with the analytical results from the continuum scale thin plate theory, they found $q=1.04$\AA.  Atomic-scale studies focusing on the computation of $\nu$ report it to vary from being negative to positive. For example, Qin et al.\cite{qin2017negative} performed MD simulations on graphene using AIREBO potential to conclude that the ripples in out-of-plane direction imply negative Poisson’s ratio $(\nu$ $\in$ $0.1$ to $-0.4)$. Jiang et al.\cite{jiang2016intrinsic} showed auxetic behaviour in graphene after $6\%$ strain from molecular static simulations, and Qin et al.\cite{qin2018origin} found a similar behaviour after $18\%$ tensile strain in the armchair direction from density functional theory. 

The values of $E, \nu$, and $q$ reported by other authors, including the method used, are summarized in table (\ref{tab:tab1}). Notice the large scatter in the values -- $E$ ranges from as small as 0.9 TPa to 7.0 TPa, while $\nu$ ranges from being negative to positive and $q$ varies from 0.75 \AA to 3.4 \AA. Consequently, there is a large uncertainty involved in the computation of effective properties of nanocomposites, and that presents a significant challenge to researchers. We believe that this large scatter is due to the inconsistent evaluation of the mechanical properties -- in most of the studies, either $q$ or $\nu$ or both are assumed, and the different properties are not found independently. To the best of our knowledge, the work by Huang et al \cite{huang2006thickness} is the only attempt at evaluating the values of $E, \nu$, and $q$ at the same time. They expanded the Brenner's potential using Taylor's series, mathematically simulated load tests, and made a comparison with continuum scale theories to obtain $E=2.69-3.81$ TPa, $\nu = 0.412$ and $q = 0.618-8.74$ \AA. These properties stand in stark contrast with those typically used for graphene.

In view of this, we revisit the problem of evaluation of continuum scale properties -- $E, \nu$, and $q$ -- of graphene in a consistent manner without invoking any assumptions on either of the three variables. All three parameters are treated as unknown and evaluated from three independent equations at the same time. The methodology developed is very general and can be adopted for any two-dimensional material. The manuscript is organized as follows. Section II elaborates on the methodology used for establishing the three independent equations. Section III provides the details of MD simulations. The results are presented in Section IV.

\section{METHODOLOGY}
Consider graphene as a thin rectangular plate of dimensions $a\times b$ in the $x-y$ plane. 
Let its thickness be $q$ along the $z$-direction. While it is easy to determine $a$ and $b$ from the geometry of graphene, accurately determining $q$ is ambiguous due to graphene being single-atomic layer thick. Should the thickness be taken as the radius of the Carbon atom or the inter-layer spacing observed in graphite? Rather than selecting $q$ in an ad-hoc manner, we present here an approach through which $q$ may be determined in a sound manner. Apart from $q$, $E$ and $\nu$ are the other unknown variables. In order to independently calculate them, three independent equations involving them are needed. In this manuscript, these three equations are obtained from (i) uniaxial tension, (ii) biaxial tension and (iii) free flexural vibrations of graphene sheet at finite temperature. 

The graphene sheet is assumed to behave like a linear isotropic material. The isotropic assumption may be justified from the independence of $E$ and $\nu$ of graphene with respect to the chiral angle \cite{zhou2013elastic}. The linearity assumption may be justified from the fact that $E \times q$, i.e. the thickness scaled elastic modulus, of graphene is undervalued by only 3\% when linear behavior is considered vis-\'a-vis nonlinear behavior. As a consequence of these assumptions, only two independent Lame's constants are sufficient to define the elastic properties of graphene. 

We now elaborate the three independent equations developed for solving the three unknowns.

\subsection{Uniaxial Tensile Test}
\begin{figure}[t]
    \centering
    \includegraphics[width=0.70\linewidth]{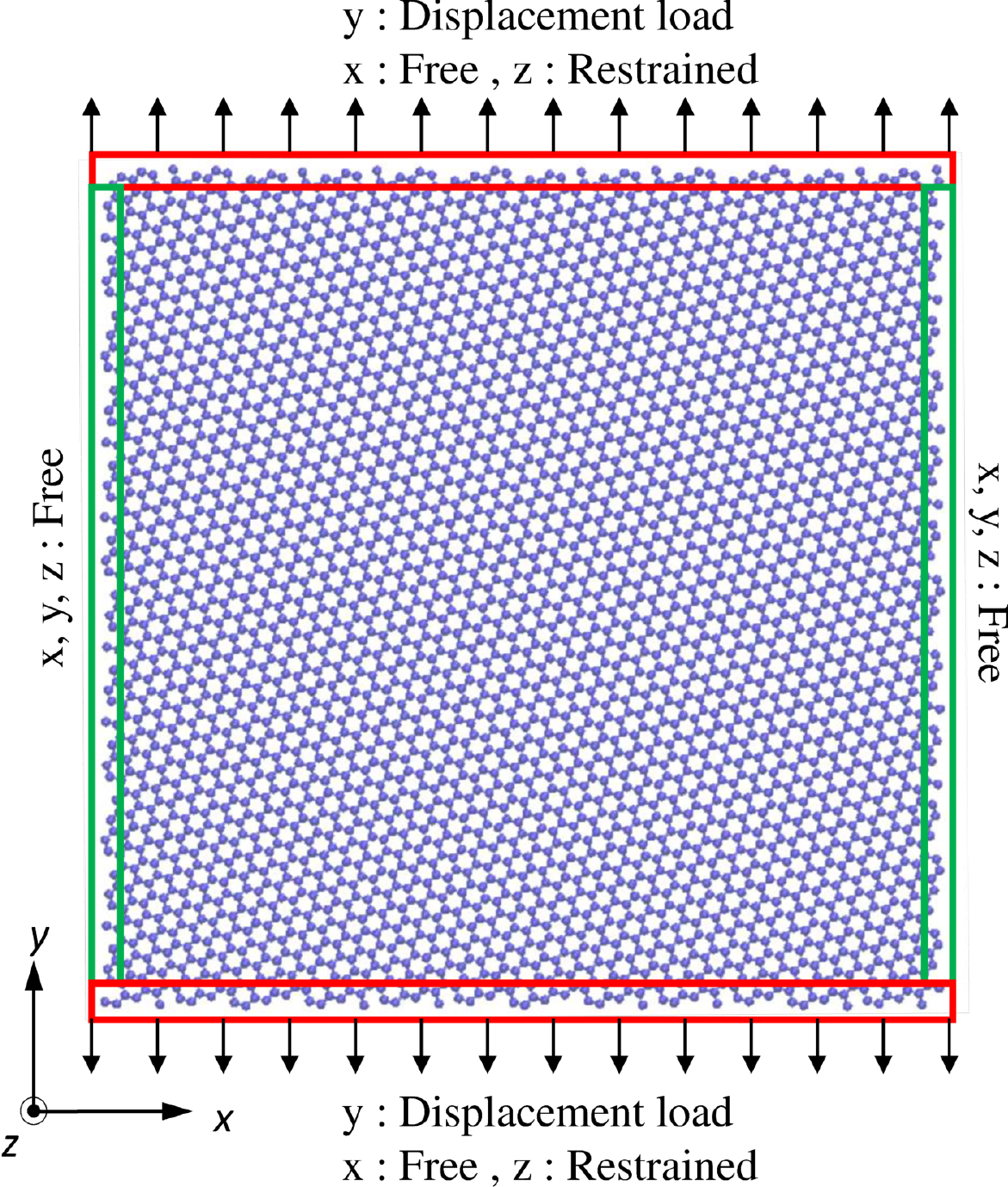} %SAURABH I AM EDITING HERE. THE FIGURE IS FINE. 
    \caption{Boundary conditions for uniaxial test}
    \label{fig:fig1}
\end{figure}
Consider a graphene sheet of dimensions described before. Let the graphene sheet be subjected to a constant strain rate based uniaxial tension in the $y$-direction. If the boundary conditions are chosen according to figure (\ref{fig:fig1}), residual stresses at the boundaries are not developed. Under these conditions, the graphene sheet behaves as a thin plate. If $U_A$ is the total strain energy and $\epsilon_{yy}$ is the strain in the $y$-direction, then:
\begin{equation}
    \dfrac{1}{V}\dfrac{\partial^2 U_A}{\partial \epsilon_{yy}^2} = E \implies \dfrac{\partial^2 U_A}{\partial \epsilon_{yy}^2} = E \times q \times a \times b,
     \label{eq:d2u-dy2}
\end{equation}
where, $V$ denotes the volume and equals: $V= q \times a \times b$. While writing equation (\ref{eq:d2u-dy2}), we have assumed that the rate of loading is so slow (quasistatic) that the work done during straining is solely equal to the increment in strain energy, and the increment in kinetic energy may be neglected. 

MD simulations readily provide the information related to strain energy in terms of po$y$-directiontential energy, and if the rate of loading is slow, the second derivative of potential energy with respect to strain is a good approximation for equation (\ref{eq:d2u-dy2}). Typically, one may expect $U_A$ to vary quadratically with $\epsilon_{yy}$ from the data of MD, and the underlying equation may be obtained from a least-squares based curve fitting. The second derivative of the equation provides an estimate of the RHS of equation (\ref{eq:d2u-dy2}).

\subsection{Biaxial Tensile Test}
\begin{figure}[t]
    \centering
    \includegraphics[width=0.80\linewidth]{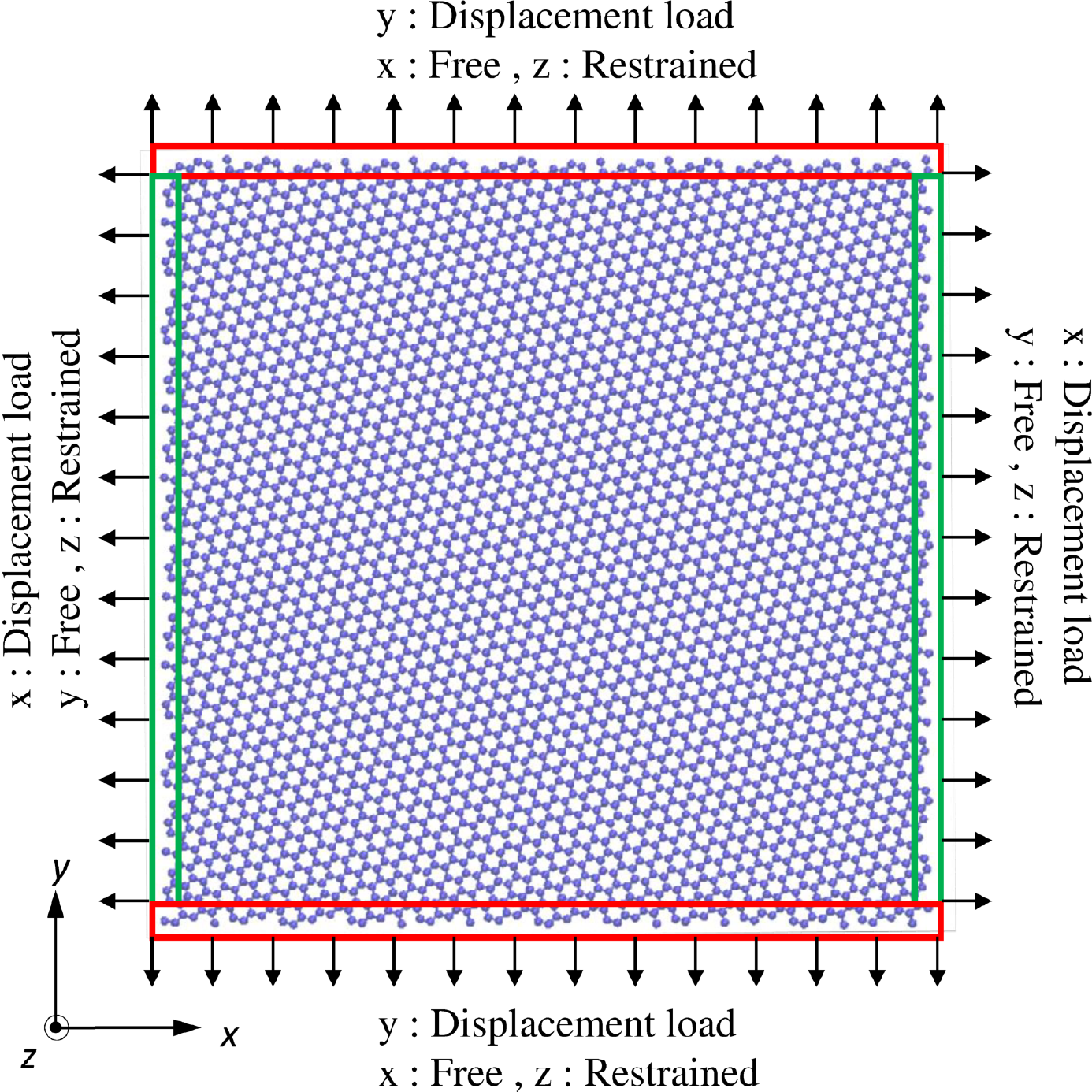}
    \caption{Boundary conditions for biaxial test}
    \label{fig:fig2}
\end{figure}
The second equation may be obtained from the biaxial tensile testing of graphene, with boundary conditions as depicted in figure (\ref{fig:fig2}). Here, the graphene sheet is loaded in both the $x$ and $y$-directions at the same constant strain rate, so that, $\epsilon_{xx} = \epsilon_{yy} = \epsilon$ at all times. In other words, an equibiaxial loading is applied. Let the normal stresses corresponding to the $x$ and $y$-direction be $\sigma_{xx}$ and $\sigma_{yy}$. Being principle stresses, the strain energy per unit volume, for this state of stress is:
\begin{equation}
    \dfrac{U_B}{V} = \dfrac{1}{2} \left( \sigma_{xx}\epsilon_{xx} + \sigma_{yy}\epsilon_{yy} \right).
    \label{eq:U_B_Vol}
\end{equation}
As graphene has been assumed to behave like a ``thin'' plate, stresses and strains may be related through the plane stress constitutive relation:
\begin{align}
   \begin{pmatrix}
        \sigma_{xx} \\
        \sigma_{yy}
   \end{pmatrix} =\frac{E}{(1-\nu^2)} \begin{bmatrix}
                        1 & \nu \\
                        \nu & 1
                   \end{bmatrix} \begin{pmatrix}
                                    \epsilon_{xx} \\
                                    \epsilon_{yy}
                                 \end{pmatrix}
                                 \label{eq:plane_stress}
\end{align}
Substituting equation (\ref{eq:plane_stress}) in equation (\ref{eq:U_B_Vol}) provides the relation between $U_B$ and the strains:
\begin{equation}
    \dfrac{U_B}{V} = \dfrac{E}{2(1-\nu^2)} \left[\epsilon_{xx}^2+\epsilon_{yy}^2+2 \nu \epsilon_{xx} \epsilon_{yy}\right]
    \label{eq:U_B_Vol_2}
\end{equation}
Equation (\ref{eq:U_B_Vol_2}) may be further simplified considering the fact that $\epsilon_{xx} = \epsilon_{yy} = \epsilon$:
\begin{equation}
\begin{array}{rcl}
    \dfrac{U_B}{V} = \dfrac{E \epsilon^2}{(1-\nu)} & \implies &  U_B = \dfrac{E q a b \epsilon^2}{(1-\nu)} \\
\end{array}
    \label{eq:U_B_Vol_3}
\end{equation}
Taking the second derivative of $U_B$ with respect to $\epsilon$, we therefore, get:
\begin{equation}
\begin{array}{rcl}
    \dfrac{\partial ^2 U_B}{\partial \epsilon^2} & = & \dfrac{2E q a b}{(1-\nu)} \\
\end{array}
    \label{eq:biaxial_equation}
\end{equation}
Like in the uniaxial case, quasistatic loading has been assumed so that the information of biaxial strain energy, $U_B$, may be obtained directly from MD simulations in terms of the potential energy.

One can directly calculate the value of Poisson's ratio from the ratio of equations (\ref{eq:biaxial_equation}) and (\ref{eq:d2u-dy2}):
\begin{equation}
\begin{array}{rcl}
    \dfrac{\partial ^2 U_B/\partial \epsilon^2}{\partial^2 U_A / \partial \epsilon_{yy}^2} & = & \dfrac{2}{(1-\nu)} \\
\end{array}
\label{eq:Poisson's ratio}
\end{equation}
By definition, Poisson's ratio may also be calculated from uniaxial tests: 
\begin{equation}
\nu = \left\langle -\dfrac{\epsilon_{xx}(t)}{\epsilon_{yy}}\right\rangle_t.
\label{eq:Poissons_ratio}
\end{equation}
Here, $\langle \ldots \rangle_t$ denotes time averages performed over an entire uniaxial simulation. Note that $\epsilon_{xx}(t)$ denotes the true strain along the $x$ direction. For the remainder of this manuscript, we use the notation $\nu_c$ when Poisson's ratio is computed using equation (\ref{eq:Poisson's ratio}) and $\nu_d$ when it is evaluated using equation (\ref{eq:Poissons_ratio}).

\subsection{Flexure Test}
With $\nu$ computed, an additional equation is necessary to compute $E$ and $q$ independently. This may be obtained from the analysis of free vibrations exhibited by a graphene sheet at a finite temperature. Our approach is similar to that given by Krishnan et al. \cite{krishnan1998young} for carbon nanotubes and Jiang et al. \cite{jiang2009young} for graphene. Consider a simply supported graphene sheet undergoing free vibrations due to it being kept at a finite temperature. At moderate temperatures ($T < 500$ K), acoustic phonon modes \cite{jiang2009young} dominate over the optical modes, and are related to the flexural vibrations of the graphene sheet. The flexural vibrations, on the other hand, are related to the flexural rigidity of the ``equivalent'' plate. Using this concept, the third equation is developed. The differences between our proposed methodology and that by Jiang et al. \cite{jiang2009young} are: (i) we use non-periodic boundaries instead of periodic boundaries, and (ii) our method incorporates surface effects owing to the omission of periodic boundaries so that the properties obtained are truly at small-scales rather than the bulk properties calculated by Jiang et. al \cite{jiang2009young}. 

Consider a continuum thin plate under plane-stress conditions, exhibiting free vibrations in the out-of-plane direction. Let the instantaneous vibration at any point ($x,y$) within the plate be denoted by $z(x,y,t)$. Neglecting shear deformations, the governing equation for $z(x,y,t)$ along with the simply supported boundary conditions may be written as:
\begin{equation}
\begin{array}{rcl}
    \dfrac{\partial^2 z(x,y,t)}{\partial t^2} + \dfrac{D}{\rho q} \Delta^2 z & = & 0, \\
    z(x=0,y,t)= 0 &, & z(x=a,y,t) = 0,\\
    z(x,y=0,t) = 0 &,& z(x,y=b,t) = 0,\\
    \left. \dfrac{\partial^2 z(x,y,t)}{\partial^2 x}\right|_{x=0,y} = 0 & , & 
    \left. \dfrac{\partial^2 z(x,y,t)}{\partial^2 x}\right|_{x=a,y} = 0,  \\
    \left. \dfrac{\partial^2 z(x,y,t)}{\partial^2 y}\right|_{x,y=0} = 0 & , &
    \left. \dfrac{\partial^2 z(x,y,t)}{\partial^2 y}\right|_{x,y=b} = 0,  \\
    \label{eq:plate-vib}
\end{array}
\end{equation}
Here, $t$ denotes the time, $D$ the flexural rigidity ($=Eq^3/12(1- \nu ^2)$), and $\rho$, the mass per unit area. 
The solution of $z(x,y,t)$ may be expressed in Fourier space as:
\begin{equation}
\begin{split}
   z(x,y,t) &=\sum_{m,n=1}^{\infty} z_{mn}(t)\\
   &= \sum_{m,n=1}^{\infty} \phi_{mn} \sin\left(\frac{m \pi x}{a}\right) \sin\left(\frac{n \pi y}{b}\right)\cos\left(w_{mn} t\right),
   \label{eq:z_mn}
\end{split}
\end{equation}
where, $\phi_{mn}$ is the Fourier coefficient corresponding to the mode ($m,n$) of frequency:
\begin{align}
     w_{mn} = \pi^2\left(\frac{m^2}{a^2}+\frac{n^2}{b^2}\right)\sqrt{\frac{D}{\rho q}} 
\end{align}

For a nanoscale plate undergoing free vibrations at a finite temperature, the total energy of vibrations and the amplitude corresponding to each mode are random variables. If these variables can be related with the thermodynamic quantities obtained from statistical mechanics, a consistent description of vibrations may be obtained. For this purpose, let us focus our attention on a specific mode $(m,n)$. Corresponding to this mode, each point on the plate vibrates periodically:
\begin{align}
    z_{mn}(t) = z_0 \cos(w_{mn}t)
    \label{eq:zmnt}
\end{align}
with a frequency $w_{mn}$ and an amplitude, $z_0$, that is dependent on the location of the point:
\begin{align}
    z_0 = \phi_{mn} \sin\left(\frac{m \pi x}{a}\right) \sin\left(\frac{n \pi y}{b}\right)
    \label{eq:z0}
\end{align}
If $E_{mn}^T$ is the total energy corresponding to this mode, then the energy of a point located at $(x,y)$ vibrating in this mode is:
\begin{equation}
\begin{array}{rcl}
E_{mn}^{xy} & = & \psi E_{mn}^T \\ 
\text{ where, } \psi & = & \dfrac{\phi_{mn}^2 \sin^2\left(\frac{m \pi x}{a}\right) \sin^2\left(\frac{n \pi y}{b}\right)}{\int \int \phi_{mn}^2 \sin^2\left(\frac{m \pi x}{a}\right) \sin^2\left(\frac{n \pi y}{b}\right) dx dy} 
\end{array}
\end{equation}
Due to the one-one mapping of $E_{mn}^{xy}$ with $E_{mn}^T$, the conditional probability density function (PDF) of the point, at any instant $t$, to lie around $z$, given it vibrates in the mode $(m,n)$ with energy  $E_{mn}^T$, may be written as:
\begin{align}
f\left(z|w_{mn},E_{mn}^T\right)= \Bigg\{\begin{split}\frac{1}{\pi\sqrt{z_0^2-z^2}} ; & \mid z \mid < z_0 \\
 0 ;    \text{Otherwise}
\end{split}
\label{eq:cond_prob}
\end{align}
We next try to find the conditional probability of the mode to have an energy $E_{mn}^T$. In order to do so, we bring in the concepts of statistical mechanics, as highlighted by Krishnan et al. \cite{krishnan1998young}. At a finite temperature, when a nanoscale plate is vibrating, the energy transport may be described in terms of phonons. Each phonon corresponding to the frequency $w_{mn}$ carries an energy given by: $E_{p} =\hbar w_{mn}$. However, the total number of phonons is not stationary at a finite temperature, and one has to probabilistically estimate the number of phonons. The probability that there are exactly $l$ phonons in the $(m,n)$ vibration mode is given by the Boltzmann's factor:
\begin{equation}
    P(l|w_{mn}) = \dfrac{\exp \left( -l\hbar w_{mn} / k_BT\right)}{1-\exp \left( \hbar w_{mn} / k_BT\right) }.
\end{equation}
To a very good approximation,
\begin{equation}
    P(l|w_{mn}) \approx \dfrac{\hbar w_{mn}}{k_BT} {\exp \left( -l \hbar w_{mn} / k_BT\right)},
\label{eq:Pl}
\end{equation}
The total energy carried by these $l$ phonons, $l \hbar w_{mn}$, is nothing but the energy, $E_{mn}^T$, of the ($m,n$) vibration mode. As the energy of a phonon is quantized, $\Delta E_{mn} = \hbar w_{mn}$, and one can rewrite equation (\ref{eq:Pl}) in terms of $E_{mn}^T$ :
\begin{equation}
    P(E_{mn}^T|w_{mn}) = \dfrac{1}{k_BT} {\exp \left( -E_{mn}^T / k_BT\right)} \Delta E_{mn},
\label{eq:PEmn1}
\end{equation}
which, in the limit of a large plate (continuum limit) becomes:
\begin{equation}
    f(E_{mn}^T|w_{mn})dE_{mn} = \dfrac{1}{k_BT} {\exp \left( -E_{mn}^T / k_BT\right)} \ d E_{mn},
\label{eq:PEmn}
\end{equation}
The conditional PDF of finding the point around $z$ while it vibrates in the $(m,n)$ mode may now be obtained by convoluting the conditional PDF shown in equation (\ref{eq:cond_prob}) with the conditional PDF shown in equation (\ref{eq:PEmn}):
\begin{equation}
\begin{array}{rcl}
    f(z|w_{mn}) & = & \int\limits_0^\infty \left[ f(z|w_{mn},E_{mn}^T) \times f(E_{mn}^T|w_{mn})\right] dE_{mn} \\
              & = & \int\limits_0^\infty \dfrac{\exp\left(-E_{mn}^T/k_BT\right)}{\sqrt{z_0^2-z^2}} dE_{mn} 
\end{array}
\label{eq:Pzmn}
\end{equation}
At any instant, $E_{mn}^T$ comprises the kinetic and potential energy of the mode. For each mode, there exists a time when the entire contribution to $E_{mn}^T$ comes from the kinetic energy. Without the loss of generality, such a situation arises when $t= \pi/2w_{mn}$, and $E_{mn}^T$ may be obtained by differentiating $z_{mn}$ shown in equation (\ref{eq:z_mn}) after multiplying with appropriate mass:
\begin{align}
\begin{split}
    E_{mn}^T & = \int\limits_0^b\int\limits_0^{a} \frac{1}{2}\rho q \left(\frac{\partial z_{mn}}{\partial t}\right)^2\bigg| _{\frac{\pi}{2w_{mn}}} \,dx\,dy\\
    & = \frac{1}{8}\rho q a b w_{mn}^2\phi_{mn}^2
    \label{eq:Emn}
\end{split}
\end{align}
Substituting the value of $z_0$ from equation (\ref{eq:z0}) and subsequently replacing $\phi_{mn}$ in terms of $E_{mn}^T$ from equation (\ref{eq:Emn}), equation (\ref{eq:Pzmn}) can be simplified to:
\begin{equation}
    f(z|w_{mn}) = \int\limits_{\lambda_{mn} z^2}^\infty \dfrac{\exp\left(-E_{mn}^T/k_BT\right)}{\sqrt{\dfrac{E_{mn}^T}{\lambda_{mn}}-z^2}} dE_{mn} 
\label{eq:Pzmn2}
\end{equation}
where,
\begin{equation}
    \lambda_{mn} = \dfrac{\rho q a b w_{mn}^2}{8\left(\sin\left(\frac{m \pi x}{a}\right) \sin\left(\frac{n \pi y}{b}\right)\right)^2}
\end{equation}
Upon integration, equation (\ref{eq:Pzmn2}) yields:
\begin{equation}
    f(z|w_{mn}) = \sqrt{\frac{\lambda_{mn}}{\pi k_B}} \exp\left(-\frac{\lambda_{mn} z^2}{k_BT}\right)
\label{eq:Pzmn3}
\end{equation}
We reiterate that equation (\ref{eq:Pzmn3}) represents the conditional PDF of finding a particle around $z$ when it vibrates in the mode $(m,n)$ with frequency $w_{mn}$. Evidently, this is a Gaussian distribution with variance, $\sigma_{mn}^2 = k_BT/(2 \lambda_{mn})$. Since in a constant temperature environment, all modes contribute independently towards determining the out-of-plane motion at a location $(x,y)$, their effect needs to be incorporated while calculating the probability of finding a point in the interval $z$ to $z+dz$. The PDF is given by: 
\begin{equation}
    f(z) = \sum\limits_{m,n=1}^\infty \sqrt{\frac{\lambda_{mn}}{\pi k_B}} \exp\left(-\frac{\lambda_{mn} z^2}{k_BT}\right),
\label{eq:Pzmn4}
\end{equation}
Notice that this is a sum of independent normal random variables, and as a result, the PDF of $z$ is also a normal random variable, with a variance given by:
\begin{equation}
\begin{array}{rcl}
    \sigma^2 & = & \sum\limits_{m,n=1}^\infty \sigma_{mn}^2 \\
    & = & \dfrac{48k_BT (1-\nu^2)}{abEq^3}\sum\limits_{m,n=1}^\infty \left(\dfrac{\sin\left(\dfrac{m \pi x}{a}\right)\sin\left(\dfrac{n \pi y}{b}\right)}{\pi^2\left(\dfrac{m^2}{a^2}+\dfrac{n^2}{b^2}\right)}\right)^2
\label{eq:thirdeq}
\end{array}
\end{equation}
This expression gives the required third equation for solving $E$ and $q$. The LHS of equation (\ref{eq:thirdeq}) may directly be obtained from MD simulations. We remind the readers that this equation has been derived from the continuum scale thin plate vibration equation (\ref{eq:plate-vib}). Due to the assumptions of continuum mechanics, no discrete particles are present within the domain, rather the matter is treated as a continuous medium. However, MD simulations contain a finite number of particles. Consequently, the graphene sheet must be long enough for it to be approximated by the continuum scale equations. Since our boundary conditions are non-periodic, our approach towards solving the unknowns is unlike that in literature, where the variance is averaged over the entire domain \cite{jiang2009young}. Instead, a symmetric region around the centre of the graphene sheet is selected, and averages are computed from the particles present there. The summation in equation (\ref{eq:thirdeq}) is kept limited to $m=n=1000$. 

\section{SIMULATION DETAILS}
\begin{figure}[h]
    \centering
    \includegraphics[width=0.7\linewidth]{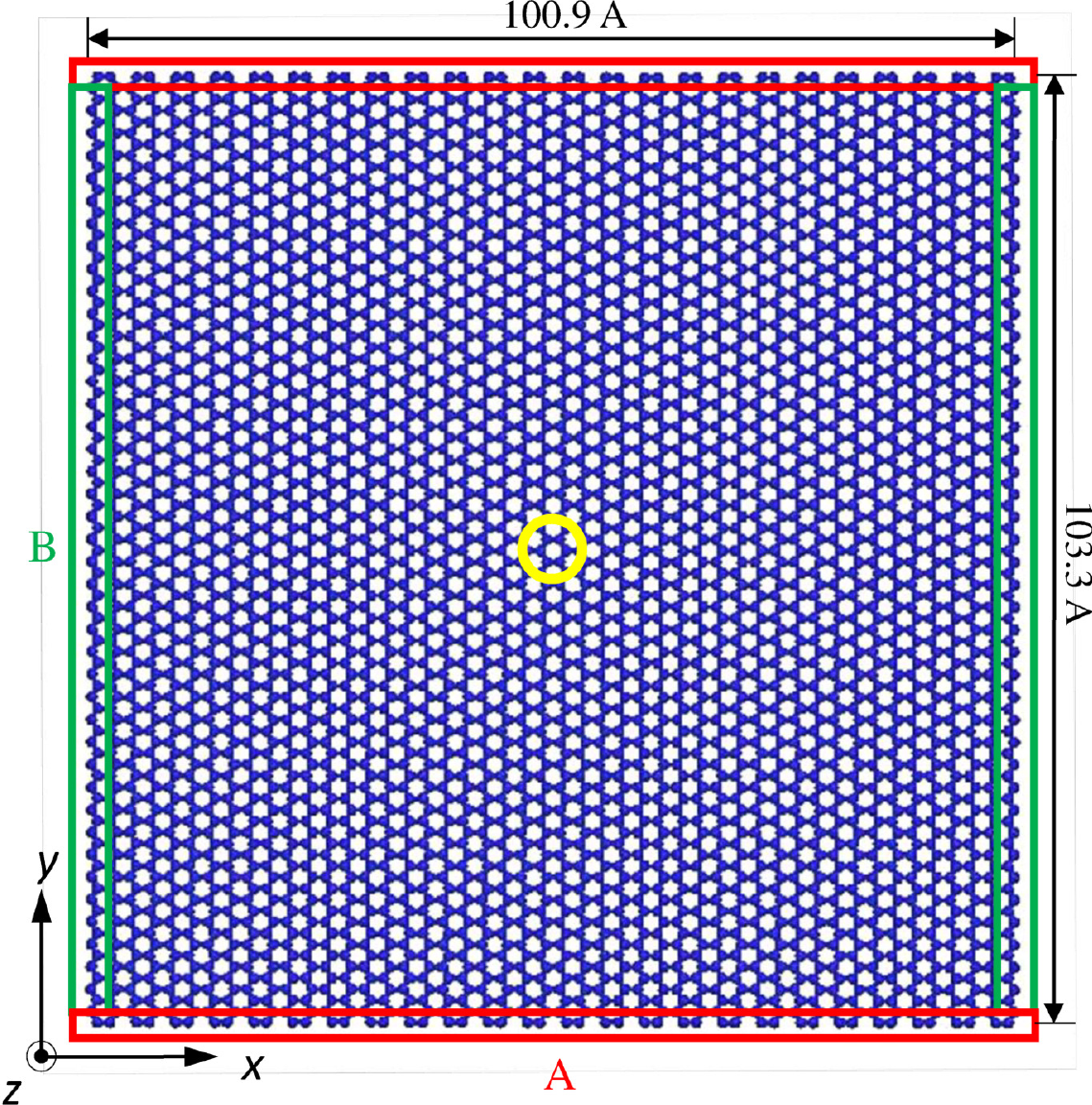}
    \caption{Geometry of plate $P_I$ -- a graphene sheet of 4080 atoms oriented along the Cartesian $x-y$ coordinate system. The top and bottom boundaries, labelled as A, have 96 atoms while the left and right boundaries, labelled as B, have 166 atoms. The region within the yellow circle has six atoms, whose instantaneous out-of-plane displacements due to thermal vibrations will be averaged for computing the displacement of the center of the plate.}
    \label{fig:p1}
\end{figure}

\begin{figure}
    \centering
    \includegraphics[width=0.70\linewidth]{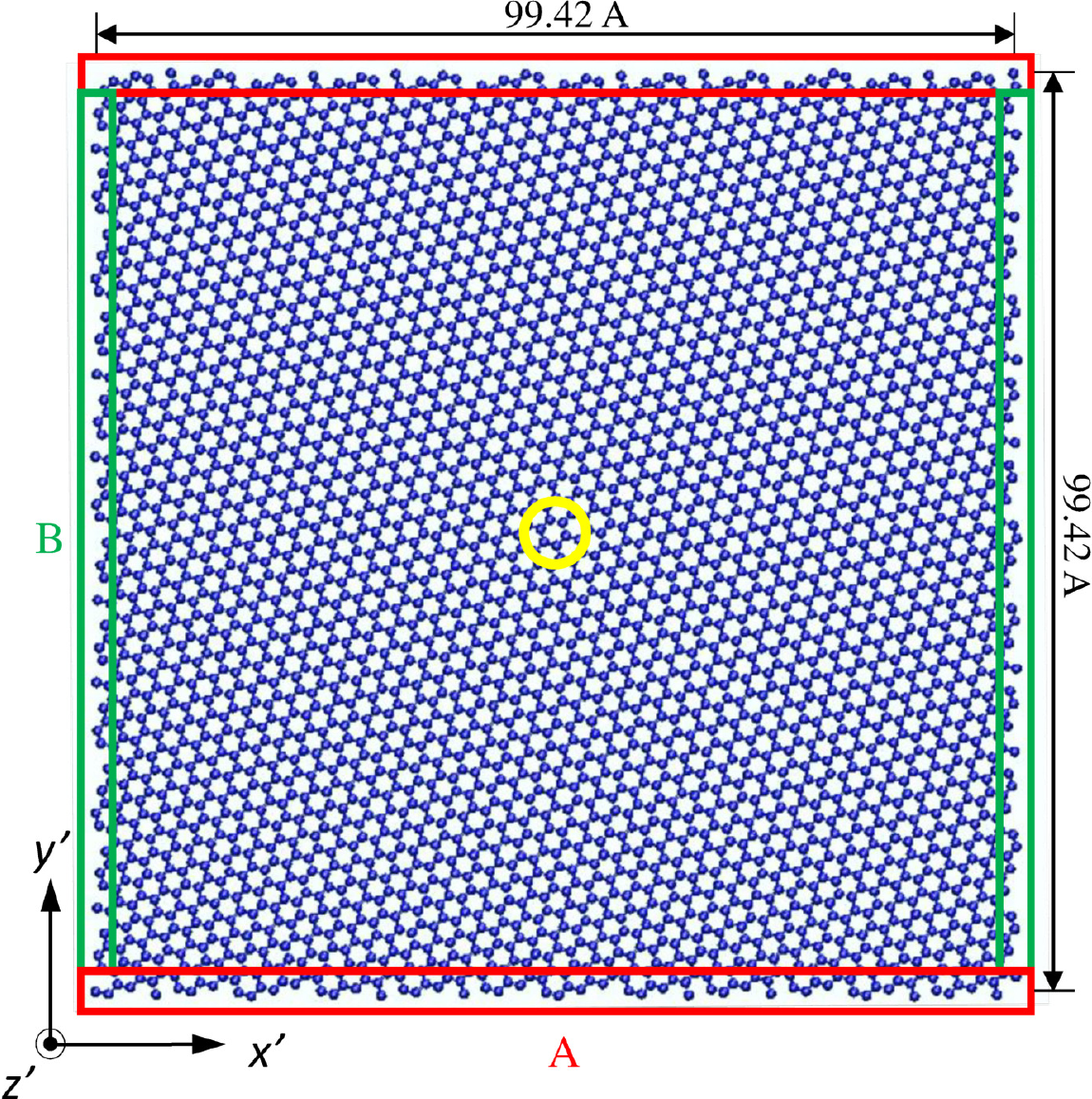}
    \caption{Geometry of plate $P_{II}$ -- a graphene sheet of 3754 atoms oriented at 45\degree with respect to the Cartesian $x-y$ coordinate system. The top and bottom boundaries, labelled as A, have 154 atoms while the left and right boundaries, labelled as B, have 148 atoms. Like in the plate $P_I$, the region within the yellow circle has six atoms.}
    \label{fig:p2}
\end{figure}

MD simulations have been performed on two plates, labelled as $P_I$ and $P_{II}$. $P_I$ comprises a $100.9 \AA \times 103.3 \AA$ graphene sheet having 4080 atoms, and oriented along the Cartesian $x-y$ coordinate system, as shown in figure (\ref{fig:p1}). Relatively larger dimensions have been chosen to ensure that the mechanical behavior of graphene sheet may be represented by the continuum scale vibration equation and to minimize the effect of size on elastic modulus, which tends to disappear when edge length is greater than $40 \AA$ \cite{jiang2009young}. The plate $P_{II}$ is a graphene sheet oriented at 45\degree to the Cartesian $x-y$ coordinate system, as shown in figure (\ref{fig:p2}). It comprises 3754 atoms, and has a dimensions of $99.42 \AA \times 99.42 \AA$. 
The reason for choosing plates with different orientations is to study the dependence of elastic properties on the loading direction. All MD simulations have been performed using free-to-use LAMMPS software \cite{Plimpton1995}. 

Interactions between the Carbon atoms of the graphene sheets have been modelled using a three body Tersoff-like potential \cite{tersoff1988empirical,tersoff1989modeling}. We choose this potential since it has seen widespread usage in the MD community for studying a variety of different problems \cite{lindsay2010optimized,suekane2008static,barreiro2008subnanometer,sircar2020simple}. In Tersoff potential, the total potential energy, $E$, is represented by:
\begin{eqnarray}
E=\sum_i{E_i}=\frac{1}{2}\sum_{i\neq{j}}\sum_j{\phi(r_{ij}}),\nonumber\\
\phi(r_{ij})=f_c(r_{ij})[f_R(r_{ij})+b_{ij}f_A(r_{ij})],
\label{eq:ab}
\end{eqnarray}
where, $E_i$ denotes the potential energy of the $i^{th}$ atom, and $\phi$ represents the interaction energy between the $i^{th}$ and $j^{th}$ atoms. The other variables of equation (\ref{eq:ab}) have the following meaning: $r_{ij}$ is the center to center distance between the atom pair $i$ and $j$, $b_{ij}$ represents the bond order function, $f_c$ is the cutoff function for ensuring the nearest-neighbor interactions, $f_R$ accounts for the repulsion between the atoms when they come close, and $f_A$ accounts for the attraction between two atoms. These functions can be represented mathematically as: 
\begin{eqnarray}
f_c(r_{ij}) = \left\{
     \begin{array}{lr}
       \text{1}&\forall  {r}_{ij}<{P}_{ij}\nonumber\\
       \frac{1}{2}-\frac{1}{2}\text{sin}(\frac{\pi}{2}\frac{r_{ij}-R_{ij}}{D_{ij}})& \forall{P}_{ij}<{r}_{ij}<{Q}_{ij}\nonumber \\
       \text{0}&\forall{r}_{ij}>{Q}_{ij}\nonumber\\
     \end{array}
   \right.
     \end{eqnarray} 
\begin{eqnarray} 
f_{R}(r_{ij})=Ae^{-\lambda_1 r_{ij}},f_A(r_{ij})=-Be^{-\lambda_2 r_{ij}},\nonumber\\  
b_{ij}=(1+\beta^n\zeta_{ij}^n)^{-\frac{1}{2n}},\nonumber\\
\zeta_{ij}=\sum_{k\neq i,j}f_C(r_{ik})g(\theta_{ijk})\text{exp}[\lambda_3^{3}(r_{ij}-r_{ik})^3],\nonumber\\
g(\theta_{ijk})=1+c^2/d^2-c^2/[d^2+(h-\text{cos} \theta_{ijk})^2)],                       
\end{eqnarray}
The cutoff function, $f_C$, is a continuous function that goes from unity to zero smoothly as the distance between two atoms vary from ${P}_{ij}={R}_{ij}-{D}_{ij}$ to ${Q}_{ij}={R}_{ij}+{D}_{ij}$. For our problem, $R_{ij}$ is chosen such that only the first neighbor shell is included. The angle between the bonds $ij$ and $ik$ is denoted by $\theta_{ijk}$. One can simulate different materials using specific values of the different parameters. In the present work, the values of the different parameters, as proposed by Lindsay and Broido \cite{lindsay2010optimized}, have been adopted.  

\subsection{Equilibrating the Graphene Sheet}
Each simulation begins with a minimization run, where the graphene sheet is relaxed using the conjugate gradient method. No boundary conditions as well as restraints are imposed on the edges of the graphene sheet at this step. Consequently, the dimensions of the sheet increase slightly, and a minimum potential energy configuration is obtained. Following minimization, researchers have traditionally equilibrated the graphene sheet in a constant pressure and temperature (NPT) ensemble \cite{lee2015effect,zhang2018negative,qin2017negative, wang2019temperature}. However, NPT equilibration poses a problem -- despite setting the pressure to zero (for avoiding any residual stress generation), the shape of the graphene sheet no longer remains rectangular. The graphene sheet becomes full of ripples, twists and wrinkles. Working with such a graphene sheet may lead to erroneous computation of the mechanical properties, and so we have used a different equilibration technique, as highlighted next. 

The graphene sheet is equilibrated for 400,000 time steps with a Langevin thermostat \cite{schneider1978molecular} instead of deterministic thermostats \cite{martyna1992nose,hoover2015ergodic,patra2015deterministic} by restraining the boundaries $A$ and $B$ (see figures (\ref{fig:p1}) and (\ref{fig:p2})) in the $z$ direction while keeping them mobile in the $x$ and $y$-directions. Note that each time step corresponds to 1 fs. This technique of equilibration maintains the rectangular shape of the graphene sheet while minimizing wrinkles, ripples and twists along with avoiding any additional thermal stresses. The residual stresses still present create strains that are significantly smaller than the strain increment imposed during the tensile tests. Being random in nature, the thermal forces, at times, induce rotation in the graphene sheet about the $z$ axis. The rotation angle, $\mathfrak{Q}$, may be calculated by taking the average of the angles by which the edges rotate about the center of the sheet.

Starting from the same post-minimization configuration, ten different equilibration runs are performed at each temperature by changing the seed of the Langevin thermostat. The state at the end of each equilibrium run serves as an initial configuration for the actual MD runs. Note that the sheets are rotated back by the angle $\mathfrak{Q}$ prior to actual MD runs. We now explain the MD simulation methodology adopted for the three tests described in the previous section.

%The equilibration runs proceed in 60 loops, where each loop comprises 20,000 time steps with every time step equalling 1 fs.
%The 20,000 time steps within a loop are split into two sets of 10,000 time steps each -- (i) in the first 10,000 time steps, the boundary particles are allowed to move in the $x$ direction while the $y$-direction is constrained, and (ii) in the last 10,000 time steps, the boundary particles are allowed to move in the $y$-direction while now the $x$ direction is constrained. 

%The residual stresses are monitored at the end of each loop, and the configuration that minimizes the residual stresses is taken as the initial configuration for further simulations.

\subsection{Uniaxial and Biaxial Tests}
Both uniaxial and biaxial tests are performed through displacement control on the configurations obtained post equilibration. Displacement control is implemented by moving the boundary atoms slowly so that the total increment in the strain energy approximately equals the increment in the potential energy. In order to allow the effect of displacement to propagate within the graphene sheet, each loading step is followed by 10,000 MD runs at constant temperature. The potential energy of the sheet is continuously monitored over these runs, and its average over these runs is taken as the potential energy corresponding to the strain.

For uniaxial tests, the atoms present in the boundary region $A$ (see figures (\ref{fig:p1}) and (\ref{fig:p2})) are displaced in the $\pm y$ direction while they remain free to move along the $x$ direction. The atoms present in the boundary region $B$ have no restrictions on their movement. A displacement rate of 0.0009 \AA/fs is chosen, which corresponds to $\approx 0.00175\%$ strain rate. 

For biaxial tests, separate MD runs are performed on the configurations obtained post equilibration. In here, the atoms in the boundary region $A$ are free to move in the $x$ direction while those in the boundary $B$ are free to move in the $y$-direction. In order to generate an equibiaxial state of loading, the displacements along the two boundaries are different: the particles in the boundary region $B$ are displaced in the $\pm x$ direction at the rate of $0.0009$ \AA/fs while those in the boundary region $A$ are displaced at the rate of $0.0009 \times (b/a)$ \AA / fs in the $\pm y$ direction. Like in the uniaxial case, the chosen value of displacement rate is such that the strain rate is $\approx 0.00175 \%$.

\subsection{Flexural Tests}
Flexural tests are performed on the configurations obtained post equilibration by imposing boundary conditions as per equation (\ref{eq:plate-vib}) -- the displacement along the $z$ direction is constrained for all boundary atoms; the displacement along the $x(y)$ direction is constrained for the atoms present in the boundary region $A(B)$. Additionally, for the atoms present in the bottom boundary region $A$ (right boundary region $B$), the displacement along $y$ ($x$) direction is also constrained. These boundaries reflect the simply supported boundary conditions used for deriving the equation (\ref{eq:thirdeq}).

The graphene sheet, subjected to the mentioned boundary conditions, undergoes free vibrations at the chosen temperature. Note that in equation (\ref{eq:thirdeq}), the variance of out-of-plane displacement is needed. Rather than working with a single particle present at the center, we choose six particles located symmetrically around the center of the sheet (shown in yellow cirular region in figures (\ref{fig:p1}) and (\ref{fig:p2}) ) in order to achieve improved convergence of variance. The flexural simulations have been performed for 20 ns with an integration time step of 1 fs. The variance reported at a specific temperature is the average over ten different sheet configurations.

\section{RESULTS}
The dimensions of graphene sheets post equilibration are different from the initial dimensions shown in figures (\ref{fig:p1}) and (\ref{fig:p2}). Due to the finite temperature effects, the sheets expand. It is on these ``expanded'' sheets we perform MD simulations, and so these dimensions serve as input to equations (\ref{eq:d2u-dy2}), (\ref{eq:biaxial_equation}) and (\ref{eq:thirdeq}) instead of the initial dimensions. The mean dimensions of the graphene sheets post equilibration at different temperatures obtained by averaging over the ten configurations are as shown in table \ref{tab:dimensions}.
\begin{table}
 \begin{tabular}{| m{2cm} | m{1.5cm} | m{2.0cm} | m{2.0cm} |} 
 \hline
 Type & $T$ (K) & $a (\AA)$ & $b(\AA)$ \\
 \hline
$P_I$ & 10 & 102.997 & 106.247 \\ 
 \hline
$P_{II}$ & 10 & 99.901 & 99.931 \\ 
 \hline
$P_I$ & 50 & 102.984 & 106.235 \\ 
 \hline 
$P_{I}$ & 100 & 102.979 & 106.221 \\ 
 \hline 
$P_{I}$ & 200 & 102.947 & 106.193 \\ 
 \hline 
$P_{I}$ & 500 & 102.955 & 106.193 \\ 
 \hline 
\end{tabular}
 \caption{Post-equilibration dimensions of graphene at different temperatures obtained by averaging over ten different configurations. While the initial dimension remains the same, as shown in figures (\ref{fig:p1}) and (\ref{fig:p2}), during minimization and equilibration, the sheets expand. The mean dimensions shown here are used for computing the mechanical properties of graphene.}
 \label{tab:dimensions}
\end{table}

\subsection{At low temperatures}
We now calculate the mechanical properties -- $E,q$ and $\nu$ -- for the two plates, $P_I$ and $P_{II}$, at 10 K. $\nu_C$ is first determined from equation (\ref{eq:Poisson's ratio}) by obtaining the dependence of $U_A$ and $U_B$ on strain, and taking their ratio. Once $\nu_C$ is determined, equation (\ref{eq:d2u-dy2}) or equation (\ref{eq:biaxial_equation}) may be used to calculate the thickness scaled elastic modulus: $E\times q$. This serves as an input to equation (\ref{eq:thirdeq}) from which $q$ can be obtained. Using these steps in the sequence described, the mechanical properties of \textit{any} two-dimensional nanoscale structure can be found.

\begin{figure}[h]
 \centering
  \includegraphics[width=0.80\linewidth]{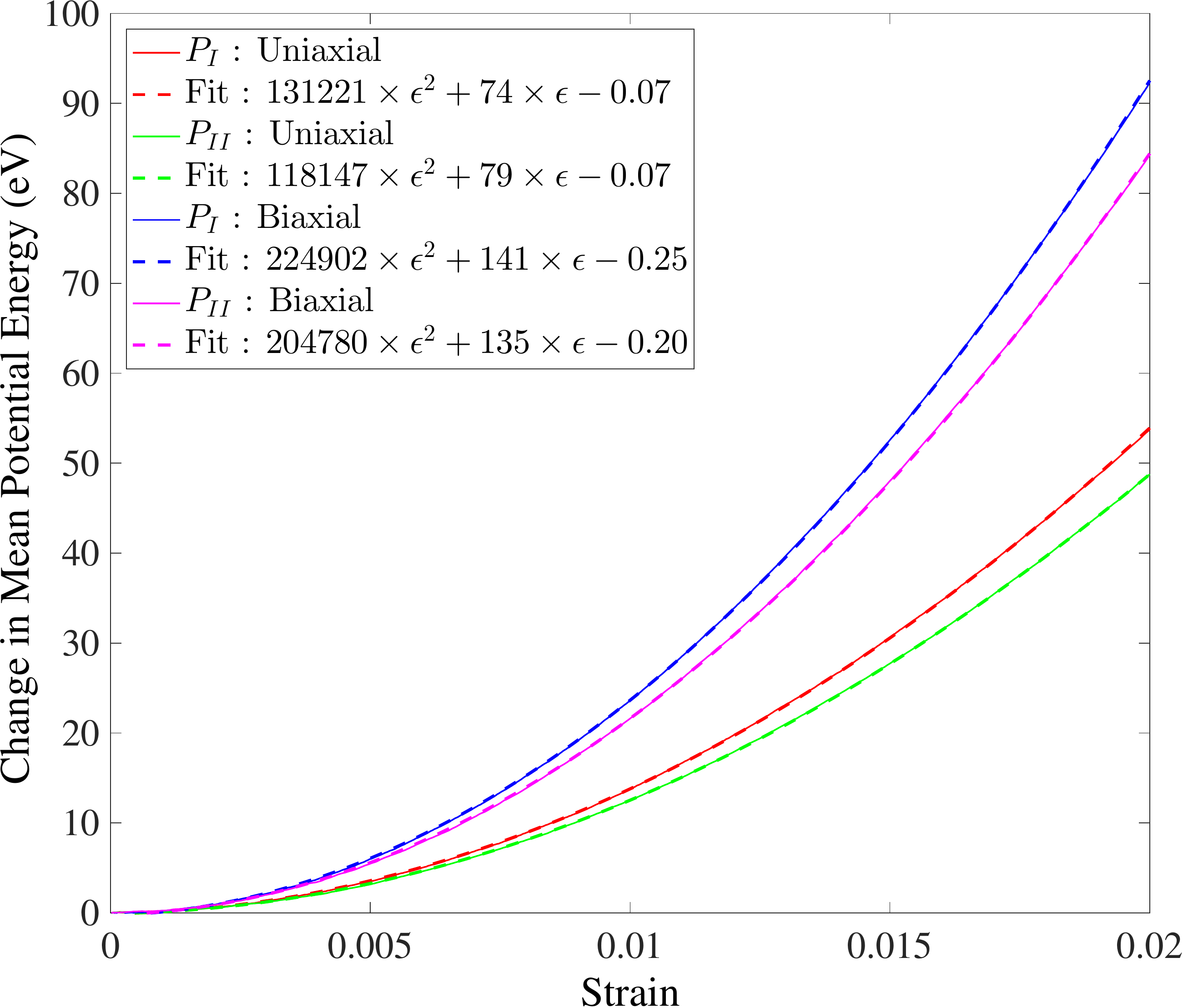}
  \caption{Change in mean strain energy of the two plates -- $P_I$ and $P_{II}$ -- under uniaxial and equibiaxial tensile loading at 10 K obtained from MD simulations (solid lines). For each case, a least square based quadratic fitting is performed, and plotted in dashed lines. There is a good agreement between the data obtained from MD simulations and the quadratic fit. Note that the mean potential energy curves are obtained by taking the average of ten configurations.}
  \label{fig:lowT-ub-comparison}
\end{figure}
Figure (\ref{fig:lowT-ub-comparison}) plots the increase in mean strain energy, $\langle \Delta U_A \rangle$ and $\langle \Delta U_B \rangle$, as the engineering strain increases to 2\%, for both the uniaxial and equibiaxial tensile tests. $\langle \ldots \rangle$ denotes an averaged quantity obtained by averaging the MD results over 10 separate configurations. Notice that the strain energy is an extensive quantity. Since, the number of atoms in $P_{II}$ are 3754 while that in $P_I$ are 4080, the strain energy increment in $P_{II}$ is marginally smaller than in $P_I$. A least squares based curve fitting is performed to obtain the quadratic dependence of the strain energies on strain. The results, shown as dashed lines in figure (\ref{fig:lowT-ub-comparison}), indicate that the increase in strain energies obtained from MD simulations can be accurately captured by second-order polynomials.

With the mathematical expressions for $U_A$ and $U_B$ determined, equation (\ref{eq:Poisson's ratio}) may be used for finding $\langle \nu_C \rangle$: $-0.1699$ for $P_I$, and $-0.1539$ for $P_{II}$. From the fundamental definition of Poisson's ratio, $\langle \nu_D \rangle = -0.1643$ and -0.1614 for plates $P_I$ and $P_{II}$, respectively. There is a marginal difference between $\langle \nu_C \rangle$ and $\langle \nu_D \rangle$: $\sim$ $3.4\%$ for $P_I$ and $\sim$ $4.6\%$ for $P_{II}$. The difference occurs since the computation of $\langle \nu_D \rangle$ involves only boundary atoms, but $\langle \nu_C \rangle$ involves all atoms of the plate, thus accounts for the true nature. 

The negative values of $\langle \nu_C \rangle$ and $\langle \nu_D \rangle$ indicate that graphene is auxetic, which is in stark contrast with graphite. The auxetic nature of graphene may be attributed to its high in-plane shear modulus vis-\'a-vis graphite, because of which the in-plane angle bending stiffness increases. Consequently, the deformation mechanism is different \cite{qin2018origin} -- there is a relatively larger increase in the lateral inline distances between the atoms since the angular deformation of the bent-angle is smaller than the axial bond deformation at lower strains. In simple terms, the atoms try to move away from each other laterally while trying to move away longitudinally, a behavior which is opposite to that of materials with positive Poisson's ratio.

The mean values of thickness scaled elastic modulus, $\langle E \times q \rangle$, can be directly determined from the uniaxial tensile tests by employing equation (\ref{eq:d2u-dy2}): $\langle E\times q \rangle =$ $384.241$ N/m and $379.223$ N/m for $P_I$ and $P_{II}$, respectively. For all practical purposes,  the small difference between the two plates ($\sim 1.3\%$) may be neglected. Our results are in agreement with the previously reported elastic stiffness of graphene: $381-385$ N/m \cite{zhou2013elastic}.

\begin{figure}[h]
  \centering
  \includegraphics[width=0.80\linewidth]{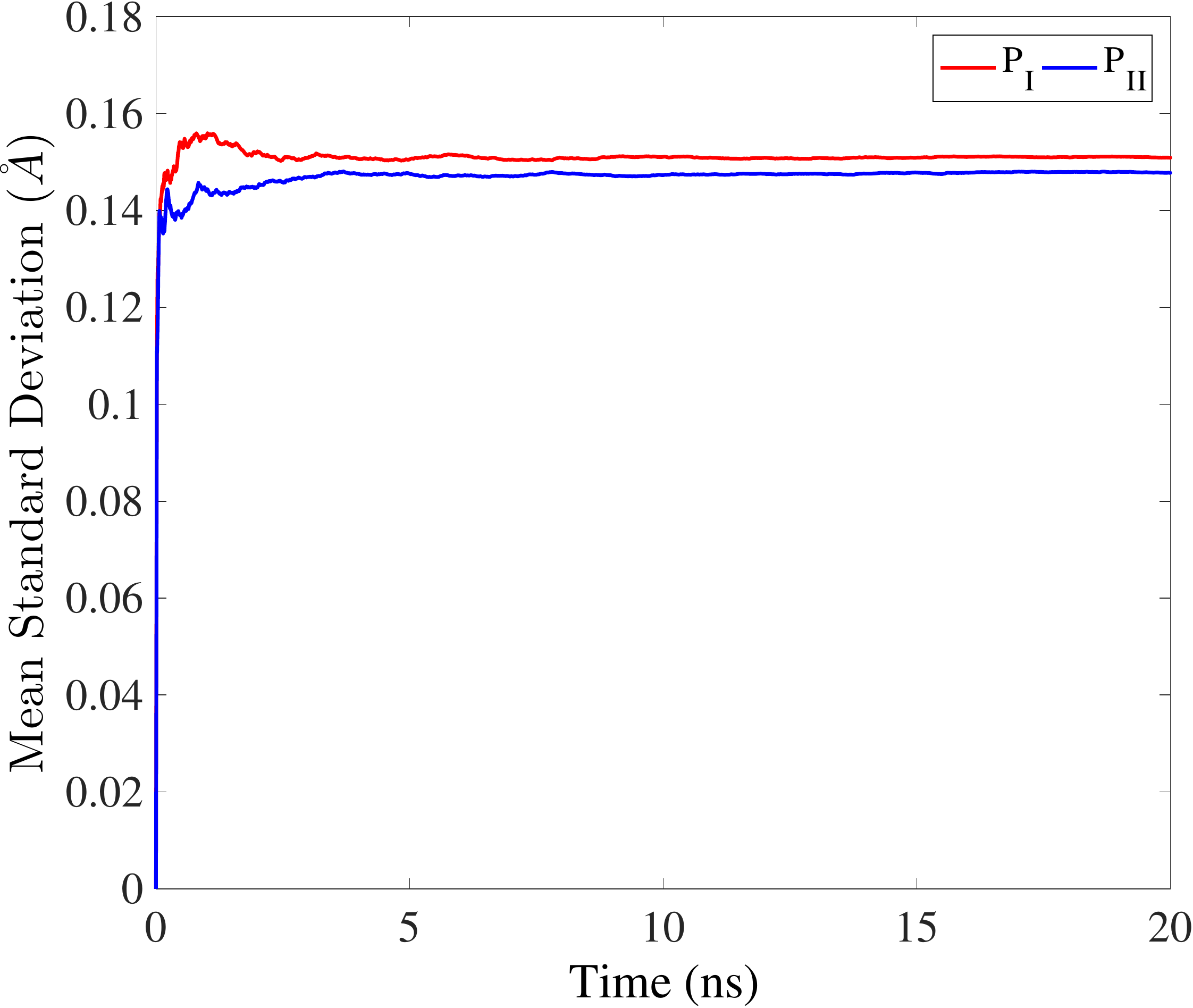}
  \caption{Convergence of mean standard deviation, $\langle \sigma \rangle$, for the two plates undergoing flexural free vibrations at 10K. The converged values of $\langle  \sigma \rangle$ for the two plates are: 0.151 \AA and 0.148 \AA for plates $P_I$ and $P_{II}$, respectively.}
  \label{fig:stdconvergence}
\end{figure}
Before analyzing the data from flexural free vibrations of graphene, the convergence of mean standard deviation, $\langle \sigma \rangle$, needs to be checked. Here, $\langle \sigma \rangle$ is obtained by averaging the $\sigma$ of equation (\ref{eq:thirdeq}) over 10 separate MD runs. The temporal evolution of $\langle \sigma \rangle$ over 20 ns for the two plates is shown in figure (\ref{fig:stdconvergence}). The converged values are found to be 0.151 \AA and 0.148 \AA for $P_I$ and $P_{II}$, respectively. As expected, $\langle \sigma \rangle$ for $P_{II}$ is smaller than $P_I$ owing to its smaller dimensions. 

With dimensions shown in table (\ref{tab:dimensions}) and $\langle E \times q \rangle$ as calculated previously, $\langle q \rangle$ for $P_I$ and $P_{II}$ equal 1.528 \AA and 1.506 \AA, respectively. The equivalent mean thicknesses are smaller than the widely used value of 3.4 \AA which denotes the inter-layer spacing of graphite \cite{sakharova2015mechanical}. There is a justification for not using 3.4 \AA as the equivalent thickness of graphene. Typically at continuum scale, thickness is inversely proportional to the flexural stiffness -- the smaller the thickness, the larger the out-of-plane deformations. Graphene, being a two-dimensional material with no ``matter'' present in the out-of-plane direction for hindering the out-of-plane deformations, shows much larger out-of-plane deformations (including wrinkles, ripples and twists) than in graphite, which comprises layers of carbon atoms that interact through van der Waals potential. The presence of van der Waals forces prevents large out-of-plane deformations in graphite vis-\'a-vis graphene. Therefore, the equivalent thickness of graphene must be smaller than that of the inter-layer spacing in graphite. 

With equivalent thicknesses known, $\langle E \rangle$ for the two plates may be obtained from the data of the $
\langle E \times q \rangle$: $ \langle E \rangle=2.515$ TPa for $P_I$ and $2.521$ TPa for $P_{II}$. These values are larger than the reported elastic modulus for graphite ($\approx 0.95$ TPa \cite{tersoff1989modeling}). Although, the structure of graphite has alternate hexagonal packed graphene sheets at nano-scale, they are in fact oriented in random way above micro-scale which reduces its elastic modulus vis-\'a-vis graphene. Further, in graphite, apart from the covalent carbon-carbon bonds, the weak inter-layer van der Waals forces start participating in the axial and lateral deformation mechanisms, which reduces the elastic modulus. The snapshot of the different variables computed at 10 K are shown in table (\ref{tab:snapshot10k}). 
\begin{table}
 \begin{tabular}{| m{1.25cm} || m{1.5cm} | m{1.0cm} | m{2.75cm} | m{1.5cm}|} 
 \hline
 \textbf{Type} & $\langle E \rangle$ (TPa) & $\langle q \rangle$ (\AA) & $\langle \nu_C \rangle$ & $\langle \nu_D \rangle$\\
 \hline \hline
$P_I$ & 2.515 & 1.528 & -0.170 & -0.164 \\ 
 \hline
 $P_{II}$ & 2.521 & 1.506 & -0.154 & -0.161 \\ 
 \hline
\end{tabular}
\caption{Snapshot of the continuum scale mechanical properties of $P_I$ and $P_{II}$. The results suggest that there is no orientation dependence in the mechanical properties of graphene sheet. }
\label{tab:snapshot10k}
\end{table}

\subsection{Temperature Dependent Properties}
\begin{figure*}
  \centering
  \includegraphics[width=0.80\linewidth]{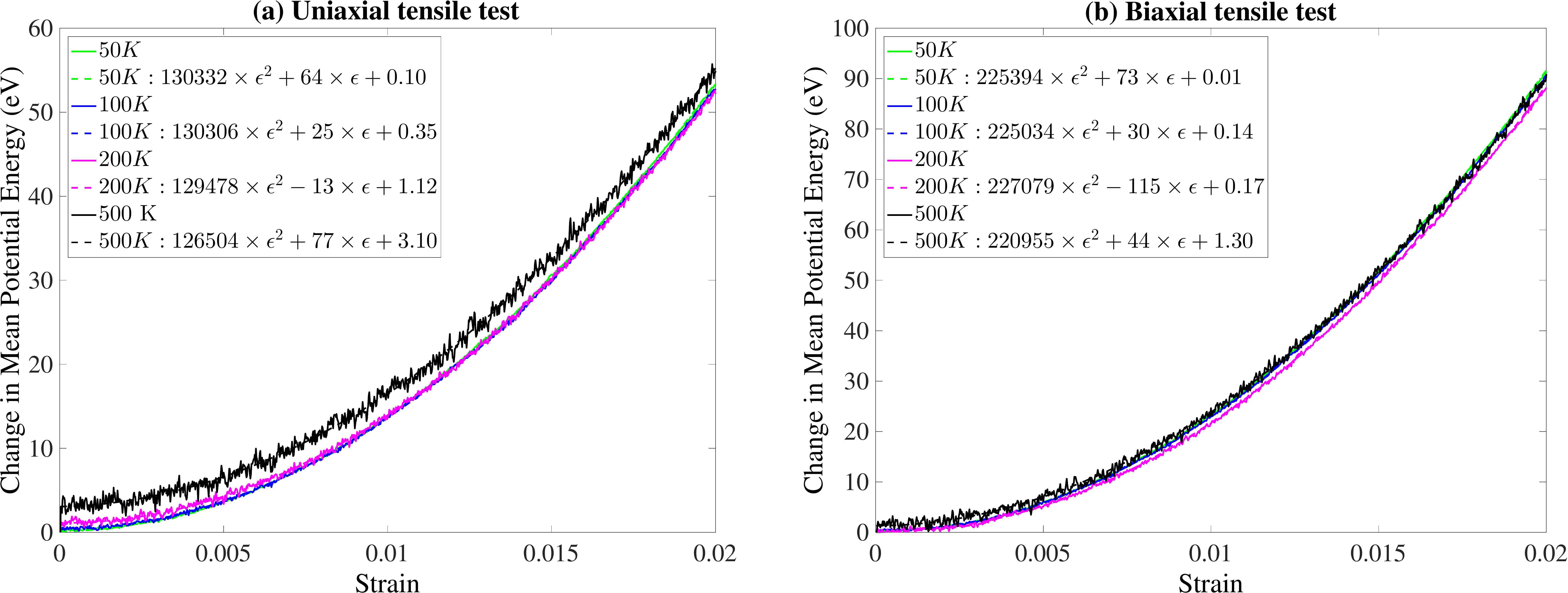}
      \caption{Increase in mean strain energy of the graphene sheet $P_I$ under: (a) uniaxial and (b) equibiaxial tensile loading, as the temperature changes from 10 K to 500 K. The average strain energy obtained from MD simulations (shown in solid lines) is subjected to a least-squares based curve fitting, which yields a parabolic dependence of strain energy on strain (shown in dashed lines). The resulting quadratic equations are indicated in the legend.}
  \label{fig:energy_multipleT}
\end{figure*}
We now study the variation of the continuum scale properties with increasing temperature. The steps for finding the three unknowns remain the same as highlighted previously. Since the values of $\langle E \rangle, \langle \nu \rangle$ and $\langle q \rangle$ are almost the same for $P_I$ and $P_{II}$ at 10 K, in this section, only the sheet $P_I$ is studied. 

The increment in mean strain energies of the sheet $P_I$ are shown in figure (\ref{fig:energy_multipleT}) (a) and (b), respectively, for uniaxial and biaxial loading, as the temperature changes from 50 K to 500 K. The solid lines correspond to averaged values obtained by averaging the MD data over 10 different configurations. For each temperature, a least-squares based fitted curve is shown in dashed lines. Like in the low temperature case, the agreement between the fitted curve and the MD data is good for all temperatures. Note that for uniaxial simulations, because of the imposed boundary conditions during equilibration runs, there is a sudden increase in potential energy once the boundaries, more specifically $z$ displacements, are relaxed. These sudden increases are more prominent at higher temperatures, as can be seen from figure (\ref{fig:energy_multipleT})(a) at 500 K. 

\begin{table}
 \begin{tabular}{| m{1.25cm} || m{1.25cm} | m{1.25cm} | m{1.25cm} | m{1.25cm}| m{1.25cm}|} 
 \hline
 $T$ (K) & 10 & 50 & 100 & 200 & 500 \\
 \hline \hline
$\langle \nu_C \rangle $ & -0.170 & -0.161 & -0.163 & -0.147 & -0.148 \\ 
 \hline
$\langle \nu_D \rangle $ & -0.164 & -0.171 & -0.173 & -0.180 & -0.191 \\ 
 \hline
\end{tabular}
\caption{Temperature dependence of $\langle \nu_C \rangle$ and $\langle \nu_D \rangle$. With increasing temperature, $\langle \nu_C \rangle$ decreases while $\langle \nu_D \rangle$ increases. This disparity occurs because in the computation of $\langle \nu_D \rangle$ only boundary atoms are taken into account. For calculation of $\langle \nu_C \rangle$, all atoms of the graphene sheet, not just the boundary particles, are considered. Boundary atoms suffer from boundary effects since atoms are present only on one side of the boundary.}
\label{tab:nu_multT}
\end{table}
Poisson's ratios, $\langle \nu_C \rangle$ and $\langle \nu_D \rangle$, are shown in table \ref{tab:nu_multT}. It is evident that graphene behaves auxetically even at temperatures as high as 500 K. With increasing temperature, while $\langle \nu_C 
\rangle$ increases, $\langle \nu_D \rangle $, on the other hand decreases. Further, the discrepancy between them increases as the temperature rises. We remind the readers that $\langle \nu_D \rangle$ only accounts for boundary particles whereas $\langle \nu_C \rangle$ accounts for all the atoms. The reason of increased disparity between the two Poisson's ratios may be attributed to the increased out-of-plane vibrations of the boundary atoms over the bulk atoms at higher temperatures. This creates a restraining effect for the movement of the boundary atoms along the lateral direction, thereby reducing $\langle \nu_D \rangle $.

\begin{table}
 \begin{tabular}{| m{1.25cm} || m{1.25cm} | m{1.25cm} | m{1.25cm} | m{1.25cm}| m{1.25cm}|} 
 \hline
 $T$ (K) & 10 & 50 & 100 & 200 & 500 \\
 \hline \hline
 $\langle E \times q \rangle$ (N/m) & 384.241 & 381.730 & 381.723 & 379.517 & 370.767 \\ 
 \hline
 $\langle \sigma \rangle $ (\AA) & 0.151 & 0.327 & 0.428 & 0.575 & 0.779 \\
 \hline
 $\langle q \rangle$ (\AA)& 1.528 & 1.588 & 1.716 & 1.817 & 2.151 \\
 \hline
 $\langle E \rangle$ (TPa) & 2.515 & 2.412 & 2.231 & 2.099 & 1.735 \\ 
 \hline
\end{tabular}
\caption{Snapshot of continuum scale properties of graphene at different temperatures. }
\label{tab:Eq_multT}
\end{table}

$\langle E \times q \rangle$ obtained from uniaxial tests and the converged values of $\langle \sigma \rangle$ obtained from flexural tests are tabulated in table \ref{tab:Eq_multT}. With increasing temperature, $\langle E \times q \rangle$ decreases. The decreased stiffness is because of the increased in-plane vibrations of the atoms around their equilibrium positions at higher temperatures. The converged values of $\langle \sigma \rangle$ indicates that both $\langle E \rangle$ and $\langle q \rangle$ are temperature dependent -- if they were temperature independent, then the ratio of $\langle \sigma \rangle$ at any two temperatures must be equal to the square root of the ratio of the two temperatures. This is reflected in the computed values of $\langle q \rangle$, which increases from 1.528\AA at 10K to 2.151\AA at 500K. On the other hand, the increase in $\langle q \rangle$ is accompanied by a decrease in $\langle E \rangle$, which is in agreement with the results reported previously \cite{shen2010temperature}.

\section{Conclusions}
The large scatter in the reported values of continuum scale elastic properties of graphene is tackled in this manuscript. The scope is kept limited to elastic modulus, $E$, which has been reported to vary between 0.912 TPa to 7 TPa, Poisson's ratio, $\nu$, which has been reported to vary from being negative to a value as large as 0.46, and effective thickness, $q$, whose value varies between 0.75 \AA to 3.41 \AA. Such a large scatter arises due to inconsistent evaluation of these properties, and making assumptions that may not be valid at atomistic scales. For example, the data from MD always provides thickness scaled elastic modulus, and to obtain the elastic modulus researchers assume the thickness. The most common assumption is to take the effective thickness of graphene to be the same as that of inter-layer spacing in graphite. However, because of the absence of weak inter-layer van der Waals forces in single-layered graphene, the assumption may not be correct. 

We combine three separate methods, which when used in tandem in MD, can provide consistent values of $E, \nu$ and $q$. The only assumption made in the present study is the validity of the continuum scale thin plate vibration equation to represent the free vibrations of a long graphene sheet. Our proposed methodology is quite general, and is suitable for any two-dimensional material. It comprises MD simulations of -- (i) uniaxial tension, (ii) equibiaxial tension, and (iii) flexural out-of-plane free vibrations on simply supported sheets. The estimate of $\nu$ and $E \times q$, obtained from uniaxial and equibiaxial tensile simulations, are subsequently used in flexural vibration simulations for computing the values of $E$ and $q$. We test our methodology on graphene, and our results suggest that -- (i) Graphene is auxetic with its Poisson's ratio increasing with increasing temperature, (ii) with increasing temperature, $\langle E \rangle$ decreases, and (iii) the effective mean thickness increases with temperature. From linear interpolation, at room temperature, $\langle \nu \rangle = -0.147, \langle q \rangle = 1.928 \AA$ and $\langle E \rangle = 1.978$ TPa. We recommend the researchers to use these values while using graphene for continuum scale experiments.

An interesting extension of this work is to understand the role played by the different potentials in determining the continuum scale properties. For example, does the different mechanical properties change if the Tersoff potential is replaced by Airebo potential in graphene. Being very general, the methodology can be adopted for finding the properties of oother two-dimensional materials such as graphyne, silicene, MoS$_2$ and Boron-Nitride sheets. There is a scope of further rationalizing the proposed methodology. Eringen's non-local elasticity theory, which reformulates continuum mechanics by accounting for the forces between the atoms and the system's length scale while constructing the constitutive equations, may be used instead of the standard continuum scale plate theories employed in the present formulation. 

\begin{acknowledgments}
Support for the research provided in part by Indian Institute of Technology Kharagpur under the grant DNI is gratefully acknowledged.
\end{acknowledgments}
%\begin{thebibliography}{50}

\bibliography{apssamp}% Produces the bibliography via BibTeX.
%Create a file called apssamp.bib

%\end{thebibliography}
\end{document}